\documentclass{Interspeech2024}

\interspeechfinaltrue

\title{Psychoacoustic Challenges Of Speech Enhancement On VoIP Platforms}

\name[affiliation={1,2}]{Joseph}{Konan}
\name[affiliation={2}]{Shikhar}{Agnihotri}
\name[affiliation={2}]{Ojas}{Bhargave}

\name[affiliation={2}]{\\ Shuo}{Han}
\name[affiliation={2}]{Yunyang}{Zeng}
\name[affiliation={2}]{Ankit}{Shah}
\name[affiliation={2}]{Bhiksha}{Raj}

\address{
  $^1$KonanAI \\
  $^2$Carnegie Mellon University }
\email{
    konan@konanai.com, jkonan@cs.cmu.edu}


\begin{document}

\maketitle

\begin{abstract}
    
    Within the ambit of VoIP (Voice over Internet Protocol) telecommunications, the complexities introduced by acoustic transformations merit rigorous analysis. This research, rooted in the exploration of proprietary sender-side denoising effects, meticulously evaluates platforms such as Google Meets and Zoom. The study draws upon the Deep Noise Suppression (DNS) 2020 dataset, ensuring a structured examination tailored to various denoising settings and receiver interfaces. A methodological novelty is introduced via Blinder-Oaxaca decomposition, traditionally an econometric tool, repurposed herein to analyze acoustic-phonetic perturbations within VoIP systems. To further ground the implications of these transformations, psychoacoustic metrics, specifically PESQ and STOI, were used to explain of perceptual quality and intelligibility. Cumulatively, the insights garnered underscore the intricate landscape of VoIP-influenced acoustic dynamics. In addition to the primary findings, a multitude of metrics are reported, extending the research purview. Moreover, out-of-domain benchmarking for both time and time-frequency domain speech enhancement models is included, thereby enhancing the depth and applicability of this inquiry. \\
\href{https://github.com/KonanAI/VoIP-DNS-Challenge}{\texttt{github.com/KonanAI/VoIP-DNS-Challenge}} 
\end{abstract}

\section{Introduction}

Voice over Internet Protocol (VoIP) has firmly established itself as an integral component of various communication paradigms, spanning corporate discussions to scholarly dialogues on global stages \cite{arora1999voice}. With its widespread adoption, pertinent issues related to audio fidelity, clarity, and preservation of acoustic nuances across multiple platforms and settings have arisen \cite{bergstra2003itu}.

In the sphere of acoustics and speech processing, the capability of VoIP to maintain speech signal integrity during real-time transmissions has been a longstanding concern \cite{rosenberg2002sip}. While challenges like packet loss, network inconsistencies, and latency have historically commanded attention \cite{bolot1993characterizing}, the contemporary integration of proprietary noise suppression techniques by industry giants necessitates a more intricate examination. Central to this discourse is understanding the impact of these advanced denoising systems on acoustics and their subsequent influences on our psychoacoustic assessments \cite{fine_grained} \cite{taploss} \cite{paaploss}.

Drawing from the vast reservoir of speech processing literature, this study establishes these goals:
\begin{enumerate}
    \item To rigorously assess modern VoIP tools, focusing on the potential acoustic anomalies arising from incorporated noise suppression algorithms \cite{haykin2008communication}.
    \item To clarify discrepancies in audio fidelity and comprehension when sound travels across diverse devices, covering both cloud-based and cellular modalities \cite{proakis2008digital}.
    \item To identify out-of-domain challenges and limitations faced by current speech enhancement models \cite{konan2023improving}.
\end{enumerate}

The scientific community's quest to unravel these dynamics extends beyond academic curiosity. Every alteration, subtle or pronounced, carries potential to significantly influence areas like voice recognition, transcription services, and auditory perception across varying scenarios \cite{davis1980comparison}. Thus, crafting a robust evaluative framework is not only relevant but crucial for the anticipated advancement of VoIP systems and their interplay with speech processing infrastructures \cite{chhetri2023speech} \cite{virtanen2012techniques}.

\begin{table*}[t]
\centering
\caption{Regression Of STOI On Acoustic Error With Interactions}\label{tab: }
\setlength{\tabcolsep}{1.5pt}
\renewcommand{\arraystretch}{1.1}
\scriptsize
\begin{tabular}{|l|c|c|c|c|c|c|c|c|c|c|c|c|c|c|c|c|c|c|c|c|c|c|c|c|c|c|r} \hline
&\textbf{X0} &\textbf{X1} &\textbf{X2} &\textbf{X3} &\textbf{X4} &\textbf{X5} &\textbf{X6} &\textbf{X7} &\textbf{X8} &\textbf{X9} &\textbf{X10} &\textbf{X11} &\textbf{X12} &\textbf{X13} &\textbf{X14} &\textbf{X15} &\textbf{X16} &\textbf{X17} &\textbf{X18} &\textbf{X19} &\textbf{X20} &\textbf{X21} &\textbf{X22} &\textbf{X23} &\textbf{X24} &\textbf{X25} \\ \hline
\cellcolor[HTML]{f3f3f3}\textbf{X} &\cellcolor[HTML]{3cb371}1.23 &0.01 &-0.01 &\cellcolor[HTML]{d0f0c0}-0.04 &0.00 &\cellcolor[HTML]{3cb371}-0.11 &\cellcolor[HTML]{3cb371}-0.03 &\cellcolor[HTML]{3cb371}0.03 &\cellcolor[HTML]{3cb371}-0.02 &\cellcolor[HTML]{77dd77}0.02 &-0.01 &\cellcolor[HTML]{3cb371}0.22 &0.01 &0.02 &\cellcolor[HTML]{3cb371}-0.26 &\cellcolor[HTML]{3cb371}0.03 &\cellcolor[HTML]{3cb371}-0.06 &\cellcolor[HTML]{3cb371}-0.23 &\cellcolor[HTML]{3cb371}0.13 &\cellcolor[HTML]{3cb371}-0.47 &-0.01 &\cellcolor[HTML]{3cb371}-0.29 &\cellcolor[HTML]{3cb371}0.52 &\cellcolor[HTML]{3cb371}0.33 &-0.05 &\cellcolor[HTML]{d0f0c0}-0.22 \\ \hline
\cellcolor[HTML]{f3f3f3}\textbf{G•X} &-0.02 &0.02 &0.05 &-0.01 &\cellcolor[HTML]{77dd77}-0.01 &-0.01 &0.02 &0.01 &-0.00 &-0.00 &0.01 &-0.02 &0.01 &\cellcolor[HTML]{77dd77}0.06 &-0.01 &0.02 &\cellcolor[HTML]{77dd77}-0.07 &0.00 &0.01 &-0.02 &0.03 &0.03 &0.21 &-0.01 &-0.02 &-0.20 \\ \hline
\cellcolor[HTML]{f3f3f3}\textbf{C•X} &\cellcolor[HTML]{3cb371}-0.09 &-0.00 &0.00 &0.00 &0.00 &\cellcolor[HTML]{3cb371}0.09 &-0.00 &-0.01 &0.01 &-0.02 &-0.02 &-0.05 &-0.02 &-0.04 &\cellcolor[HTML]{3cb371}0.15 &0.01 &-0.02 &0.06 &-0.11 &\cellcolor[HTML]{3cb371}0.46 &-0.03 &\cellcolor[HTML]{77dd77}0.16 &\cellcolor[HTML]{3cb371}-0.55 &-0.06 &-0.03 &0.11 \\ \hline
\cellcolor[HTML]{f3f3f3}\textbf{D•X} &\cellcolor[HTML]{77dd77}0.08 &\cellcolor[HTML]{d0f0c0}0.04 &\cellcolor[HTML]{77dd77}-0.08 &\cellcolor[HTML]{77dd77}0.08 &\cellcolor[HTML]{d0f0c0}0.01 &-0.02 &-0.02 &\cellcolor[HTML]{3cb371}-0.08 &0.01 &\cellcolor[HTML]{3cb371}-0.05 &\cellcolor[HTML]{3cb371}-0.08 &0.01 &\cellcolor[HTML]{d0f0c0}-0.04 &-0.04 &\cellcolor[HTML]{d0f0c0}0.05 &0.00 &0.05 &\cellcolor[HTML]{3cb371}0.20 &0.01 &\cellcolor[HTML]{77dd77}0.33 &\cellcolor[HTML]{3cb371}-0.39 &-0.04 &\cellcolor[HTML]{3cb371}-1.00 &\cellcolor[HTML]{d0f0c0}0.18 &0.09 &\cellcolor[HTML]{3cb371}0.71 \\ \hline
\cellcolor[HTML]{f3f3f3}\textbf{G•C•X} &0.02 &-0.03 &-0.07 &0.02 &0.01 &0.02 &0.01 &-0.01 &0.01 &-0.02 &0.01 &\cellcolor[HTML]{d0f0c0}0.09 &0.00 &-0.04 &-0.04 &0.00 &0.06 &0.04 &0.08 &-0.02 &-0.03 &-0.15 &-0.31 &-0.04 &0.08 &0.26 \\ \hline
\cellcolor[HTML]{f3f3f3}\textbf{G•D•X} &\cellcolor[HTML]{3cb371}-0.13 &0.02 &\cellcolor[HTML]{3cb371}-0.18 &-0.02 &\cellcolor[HTML]{3cb371}0.05 &\cellcolor[HTML]{d0f0c0}0.04 &\cellcolor[HTML]{d0f0c0}-0.04 &\cellcolor[HTML]{d0f0c0}0.05 &-0.01 &0.01 &0.01 &-0.02 &0.01 &-0.06 &\cellcolor[HTML]{d0f0c0}-0.05 &0.00 &\cellcolor[HTML]{77dd77}0.11 &0.06 &-0.08 &-0.14 &0.05 &0.04 &-0.12 &-0.08 &0.04 &0.18 \\ \hline
\cellcolor[HTML]{f3f3f3}\textbf{C•D•X} &\cellcolor[HTML]{3cb371}-0.12 &0.04 &0.01 &-0.07 &0.02 &-0.01 &-0.04 &0.01 &-0.01 &\cellcolor[HTML]{3cb371}0.11 &\cellcolor[HTML]{3cb371}0.11 &\cellcolor[HTML]{77dd77}0.18 &0.03 &0.05 &\cellcolor[HTML]{3cb371}-0.38 &0.03 &0.01 &\cellcolor[HTML]{77dd77}-0.18 &0.02 &0.23 &\cellcolor[HTML]{77dd77}0.41 &0.03 &0.45 &\cellcolor[HTML]{d0f0c0}-0.30 &0.00 &\cellcolor[HTML]{77dd77}-0.84 \\ \hline
\cellcolor[HTML]{f3f3f3}\textbf{G•C•D•X} &\cellcolor[HTML]{77dd77}0.13 &\cellcolor[HTML]{d0f0c0}-0.09 &\cellcolor[HTML]{77dd77}0.20 &0.05 &\cellcolor[HTML]{3cb371}-0.06 &-0.03 &\cellcolor[HTML]{3cb371}0.11 &\cellcolor[HTML]{d0f0c0}-0.08 &-0.00 &-0.04 &-0.05 &\cellcolor[HTML]{77dd77}-0.27 &0.02 &\cellcolor[HTML]{d0f0c0}0.12 &0.05 &0.01 &\cellcolor[HTML]{77dd77}-0.19 &0.03 &0.00 &\cellcolor[HTML]{77dd77}-0.64 &-0.20 &0.16 &0.46 &0.10 &-0.11 &0.43 \\
\hline
\end{tabular}
\end{table*}
\begin{table}[t]\centering
\scriptsize
\begin{tabular}{|lr|lr|lr|}\hline
\cellcolor[HTML]{3cb371} & $0.00 < P \leq 0.01$ &\cellcolor[HTML]{77dd77} & $0.01 < P \leq 0.05$ &\cellcolor[HTML]{d0f0c0} & $0.05 < P \leq 0.10$ \\
\hline
\end{tabular}
\end{table}

\begin{table}[t]
\centering
\caption{Blinder–Oaxaca Decomposition of STOI}\label{tab:table1}
\setlength{\tabcolsep}{3.5pt}
\renewcommand{\arraystretch}{1}
\scriptsize
\begin{tabular}{|l|c|c|c|c|c|c|c|c|} 
\hline
        & G & C & D & Endowment & Coefficient & Interaction & Collective \\
\hline
1       & 0 & 0 & 0 & -0.366 & 0.000 & 0.000 & -0.366 \\
G       & 1 & 0 & 0 & -0.364 & 0.062 & 0.050 & -0.252 \\
C       & 0 & 1 & 0 & -0.121 & 0.066 & 0.057 & 0.002  \\
D       & 0 & 0 & 1 & -0.339 & 0.018 & -0.040 & -0.361 \\
G•C    & 1 & 1 & 0 & -0.286 & 0.093 & 0.074 & -0.119 \\
G•D    & 1 & 0 & 1 & -0.460 & 0.043 & 0.007 & -0.409 \\
C•D    & 0 & 1 & 1 & -0.245 & 0.043 & 0.007 & -0.196 \\
G•C•D  & 1 & 1 & 1 & -0.386 & 0.075 & 0.043 & -0.269 \\
\hline
\end{tabular}
\end{table}

\section{Dataset and Experiment Design}

The cornerstone of this investigation rests upon the utilization of the Deep Noise Suppression (DNS) 2020 dataset. This dataset, recognized for its robustness within the domain, encompasses a set of 150 test audio samples, each with a duration of ten seconds. In addition, 1200 training audio samples are synthesized, each spanning thirty seconds \cite{reddy2020interspeech}. This structured compilation offers both depth and breadth for analysis, reminiscent of classic controlled experiment design \cite{campbell2015experimental}.

Our research paradigm is oriented around three indicator variables. The first is the selection of platform, wherein Google Meets (G = 1) and Zoom (G = 0) have been chosen. The second pertains to the sender-side denoising configuration within these platforms. For the sake of terminological uniformity across the platforms, we have streamlined the classifications to "on" (D = 1) and "off" (D = 0) regardless of native platform-specific designations. The third variable, and arguably of substantial import, focuses on the receiving interface, either the platform’s remote cloud recording (C = 1) or the experiment’s physical cellular phone recording (C = 0).

Our procedure involved each audio segment from the dataset being transmitted using a virtual microphone. This was interfaced with a NUC10i5FNH computer. This equipment configuration ensures an optimal connectivity experience, with transmission data rates surpassing 300Mbps \cite{nuc_user_guide}. Synchronously, with the audio's transmission, a cloud recording was initialized on the respective platform, with an ensuing session on an A13 5G mobile apparatus via a MixPre6-II audio interface \cite{samsung_galaxy_a13_manual} \cite{mixpre6_ii_manual}. This methodological schema was steadfastly maintained across platforms and denoising configurations.

Notwithstanding the rigorous approach, certain inherent limitations pervade. The VoIP-DNS-Tiny dataset, while admirably congruent with the research objectives, exhibits constraints. These include a certain uniformity in network configurations, and a lack of variability in sender-receiver locales and devices. Furthermore, the dataset, while comprehensive, may be somewhat strained under rigorous training procedures. An acknowledgment of these limitations not only reinforces the integrity of this study but also underscores the avenues for future research aimed at refining our domain robustness.

\section{VoIP Determinants Of Psychoacoustics}

Within the comprehensive realm of VoIP telecommunications, we stand at an intersection of traditional understanding and the pressing need to delve into the intricacies of acoustic transformations, especially given the contemporary sophistication of transmission algorithms \cite{william2002voip}. Historically, we have leveraged traditional metrics, which while robust, may not illuminate the full gamut of subtleties introduced by the modern-day VoIP mechanisms \cite{rosenberg2002sip}. Consequently, this exposition directs its focus towards an in-depth assessment employing PESQ \cite{rix2002perceptual} and STOI \cite{taal2011algorithm}, two metrics bearing significant psychoacoustic merit. These particular metrics, when viewed within the broader constellation of acoustic parameters, allow us to draw more granulated insights into the modulation patterns of speech signals within VoIP systems.

This investigation diverges from convention by eschewing traditional recognition paradigms. Instead, it casts its net over analytical frameworks, prominently featuring the Blinder–Oaxaca decomposition \cite{oaxaca1973male} \cite{blinder1973wage}—a tool traditionally entrenched in the domain of econometrics. This analytical pivot seeks to accentuate the contrasts present between target and VoIP-altered acoustics. This renders a robust, data-backed portrayal of the shifts that transpire end-to-end over VoIP architectures \cite{flanagan2012voip}.

\subsection{Analytic Methodology}

Let \( Y_{\text{PESQ}} \)\cite{rix2002perceptual} and \( Y_{\text{STOI}} \)\cite{taal2011algorithm} denote perceptual quality and intelligibility measures. Predictors \( \{X_i\} \), where \( i \in [1,25] \), are acoustic features. For a detailed and nuanced reading of each acoustic, please refer to openSMILE. \cite{egemaps}
\begin{table}[h]
\centering
\caption{Acoustic Speech Characteristics\textbf{}}
\small 
\setlength{\tabcolsep}{2.5pt} 
\begin{tabular}{@{}|ll|ll|@{}}
\hline
 & Description & & Description \\
\hline
\(X_0\)  & Intercept (Constant 1)& \(X_{13}\) & shimmerLocaldB \\
\(X_1\)  & Loudness & \(X_{14}\) & HNRdBACF \\
\(X_2\)  & alphaRatio & \(X_{15}\) & logRelF0-H1-H2 \\
\(X_3\)  & hammarbergIndex & \(X_{16}\) & logRelF0-H1-A3 \\
\(X_4\)  & slope0-500 & \(X_{17}\) & F1frequency \\
\(X_5\)  & slope500-1500 & \(X_{18}\) & F1bandwidth \\
\(X_6\)  & spectralFlux & \(X_{19}\) & F1amplitudeLogRelF0 \\
\(X_7\)  & mfcc1 & \(X_{20}\) & F2frequency \\
\(X_8\)  & mfcc2 & \(X_{21}\) & F2bandwidth \\
\(X_9\)  & mfcc3 & \(X_{22}\) & F2amplitudeLogRelF0 \\
\(X_{10}\) & mfcc4 & \(X_{23}\) & F3frequency \\
\(X_{11}\) & F0semitoneFrom27.5Hz & \(X_{24}\) & F3bandwidth \\
\(X_{12}\) & jitterLocal & \(X_{25}\) & F3amplitudeLogRelF0 \\
\hline
\end{tabular}
\end{table}

\begin{table*}[t]
\centering
\caption{Regression Of PESQ On Acoustic Error With Interactions}\label{tab: }
\setlength{\tabcolsep}{1.5pt}
\renewcommand{\arraystretch}{1.1}
\scriptsize
\begin{tabular}{|l|c|c|c|c|c|c|c|c|c|c|c|c|c|c|c|c|c|c|c|c|c|c|c|c|c|c|r} \hline
&\textbf{X0} &\textbf{X1} &\textbf{X2} &\textbf{X3} &\textbf{X4} &\textbf{X5} &\textbf{X6} &\textbf{X7} &\textbf{X8} &\textbf{X9} &\textbf{X10} &\textbf{X11} &\textbf{X12} &\textbf{X13} &\textbf{X14} &\textbf{X15} &\textbf{X16} &\textbf{X17} &\textbf{X18} &\textbf{X19} &\textbf{X20} &\textbf{X21} &\textbf{X22} &\textbf{X23} &\textbf{X24} &\textbf{X25} \\ \hline
\cellcolor[HTML]{f3f3f3}\textbf{X} &\cellcolor[HTML]{3cb371}4.73 &0.10 &0.08 &\cellcolor[HTML]{3cb371}-0.51 &\cellcolor[HTML]{d0f0c0}0.03 &\cellcolor[HTML]{3cb371}-0.63 &-0.06 &0.10 &\cellcolor[HTML]{3cb371}-0.13 &0.03 &\cellcolor[HTML]{77dd77}-0.13 &\cellcolor[HTML]{3cb371}0.58 &0.03 &\cellcolor[HTML]{77dd77}0.30 &\cellcolor[HTML]{3cb371}-1.30 &\cellcolor[HTML]{3cb371}0.29 &\cellcolor[HTML]{3cb371}-0.72 &\cellcolor[HTML]{3cb371}-1.33 &\cellcolor[HTML]{3cb371}1.48 &-0.47 &0.38 &\cellcolor[HTML]{3cb371}-2.08 &\cellcolor[HTML]{77dd77}2.10 &\cellcolor[HTML]{3cb371}1.89 &-0.30 &\cellcolor[HTML]{3cb371}-2.13 \\ \hline
\cellcolor[HTML]{f3f3f3}\textbf{G•X} &0.24 &-0.11 &0.27 &-0.18 &-0.02 &-0.13 &0.13 &-0.11 &0.06 &-0.07 &-0.01 &0.09 &0.09 &0.15 &-0.06 &0.05 &\cellcolor[HTML]{77dd77}-0.39 &0.02 &-0.37 &-0.41 &-0.01 &0.60 &0.65 &0.39 &-0.35 &-0.43 \\ \hline
\cellcolor[HTML]{f3f3f3}\textbf{C•X} &\cellcolor[HTML]{77dd77}0.37 &\cellcolor[HTML]{3cb371}-0.71 &\cellcolor[HTML]{d0f0c0}0.34 &0.13 &0.05 &\cellcolor[HTML]{d0f0c0}0.19 &\cellcolor[HTML]{3cb371}0.47 &\cellcolor[HTML]{3cb371}-0.49 &\cellcolor[HTML]{3cb371}-0.23 &\cellcolor[HTML]{3cb371}-0.25 &0.00 &\cellcolor[HTML]{d0f0c0}0.37 &0.02 &\cellcolor[HTML]{77dd77}-0.45 &\cellcolor[HTML]{77dd77}0.30 &-0.04 &0.15 &0.01 &\cellcolor[HTML]{77dd77}-0.82 &0.87 &\cellcolor[HTML]{d0f0c0}-1.29 &0.36 &-1.71 &\cellcolor[HTML]{d0f0c0}1.32 &-0.19 &0.76 \\ \hline
\cellcolor[HTML]{f3f3f3}\textbf{D•X} &\cellcolor[HTML]{77dd77}0.46 &\cellcolor[HTML]{77dd77}0.29 &\cellcolor[HTML]{3cb371}-1.31 &\cellcolor[HTML]{3cb371}1.35 &\cellcolor[HTML]{d0f0c0}0.05 &\cellcolor[HTML]{77dd77}-0.20 &\cellcolor[HTML]{77dd77}-0.21 &\cellcolor[HTML]{3cb371}-0.53 &\cellcolor[HTML]{3cb371}0.27 &\cellcolor[HTML]{77dd77}-0.23 &\cellcolor[HTML]{3cb371}-0.42 &0.18 &-0.15 &-0.14 &0.16 &0.09 &0.19 &\cellcolor[HTML]{d0f0c0}0.52 &-0.48 &1.01 &-0.69 &-0.43 &\cellcolor[HTML]{3cb371}-3.52 &0.38 &0.44 &\cellcolor[HTML]{77dd77}2.62 \\ \hline
\cellcolor[HTML]{f3f3f3}\textbf{G•C•X} &0.21 &\cellcolor[HTML]{3cb371}0.50 &-0.27 &0.10 &0.02 &0.23 &\cellcolor[HTML]{3cb371}-0.43 &-0.16 &-0.07 &-0.22 &-0.05 &-0.18 &-0.25 &-0.10 &0.15 &0.04 &\cellcolor[HTML]{77dd77}0.59 &-0.20 &0.02 &1.49 &0.96 &-0.53 &-2.07 &-0.78 &0.10 &0.71 \\ \hline
\cellcolor[HTML]{f3f3f3}\textbf{G•D•X} &\cellcolor[HTML]{3cb371}-0.94 &-0.25 &0.37 &-0.24 &-0.01 &0.21 &0.20 &\cellcolor[HTML]{77dd77}0.43 &0.04 &0.16 &-0.06 &\cellcolor[HTML]{77dd77}-0.75 &0.14 &-0.14 &-0.29 &0.01 &0.34 &0.29 &0.49 &-1.90 &-0.63 &0.42 &\cellcolor[HTML]{77dd77}4.23 &-0.81 &0.36 &-2.04 \\ \hline
\cellcolor[HTML]{f3f3f3}\textbf{C•D•X} &-0.23 &-0.20 &\cellcolor[HTML]{3cb371}1.24 &\cellcolor[HTML]{3cb371}-1.77 &-0.01 &0.13 &0.06 &\cellcolor[HTML]{3cb371}0.82 &\cellcolor[HTML]{3cb371}-0.47 &\cellcolor[HTML]{3cb371}-0.58 &\cellcolor[HTML]{77dd77}0.42 &\cellcolor[HTML]{77dd77}-0.98 &0.18 &0.18 &0.43 &0.05 &\cellcolor[HTML]{d0f0c0}-0.56 &0.61 &-0.56 &\cellcolor[HTML]{d0f0c0}-2.29 &0.51 &\cellcolor[HTML]{3cb371}1.60 &\cellcolor[HTML]{77dd77}4.78 &-1.52 &-0.04 &-1.86 \\ \hline
\cellcolor[HTML]{f3f3f3}\textbf{G•C•D•X} &0.61 &\cellcolor[HTML]{d0f0c0}0.57 &-0.24 &0.39 &-0.04 &\cellcolor[HTML]{3cb371}-0.80 &-0.10 &0.17 &0.08 &-0.44 &\cellcolor[HTML]{d0f0c0}0.43 &0.96 &0.18 &\cellcolor[HTML]{d0f0c0}-0.72 &0.14 &-0.15 &-0.55 &-0.46 &-0.30 &2.91 &-0.22 &-0.44 &-3.07 &\cellcolor[HTML]{d0f0c0}2.61 &-0.72 &0.34 \\
\hline
\end{tabular}
\end{table*}
\begin{table}[t]\centering
\scriptsize
\begin{tabular}{|lr|lr|lr|}\hline
\cellcolor[HTML]{3cb371} & $0.00 < P \leq 0.01$ &\cellcolor[HTML]{77dd77} & $0.01 < P \leq 0.05$ &\cellcolor[HTML]{d0f0c0} & $0.05 < P \leq 0.10$ \\
\hline
\end{tabular}
\end{table}
\begin{table}[t]
\centering
\caption{Blinder–Oaxaca Decomposition of PESQ}\label{tab:table1}
\setlength{\tabcolsep}{3.5pt}
\renewcommand{\arraystretch}{1}
\scriptsize
\begin{tabular}{|l|c|c|c|c|c|c|c|c|}  
\hline
        & G & C & D  & Endowment & Coefficient & Interaction & Collective \\ 
\hline
1      & 0 & 0 & 0 & -1.872 & 0.000 & 0.000 & -1.872 \\
G      & 1 & 0 & 0 & -1.800 & -0.577 & -0.055 & -2.432 \\
C      & 0 & 1 & 0 & -0.798 & -0.827 & -0.556 & -2.181  \\
D      & 0 & 0 & 1 & -1.625 & -0.750 & -0.402 & -2.777 \\
G•C    & 1 & 1 & 0 & -1.501 & -0.815 & -0.354 & -2.669 \\
G•D    & 1 & 0 & 1 & -2.188 & -0.754 & -0.273 & -3.216 \\
C•D    & 0 & 1 & 1 & -1.365 & -1.030 & -0.641 & -3.037 \\
G•C•D  & 1 & 1 & 1 & -1.934 & -0.969 & -0.480 & -3.382 \\
\hline
\end{tabular}
\end{table}

Each feature refers to a distinct speech characteristic 
using $L_1$ norm to evaluate precision. The intercept is defined \( X_0 = 1 \).
\noindent
We have three binary indicators:
\begin{enumerate}
    \item \( G \): 1 for Google Meets, otherwise 0 for Zoom.
    \item \( C \): 1 for Cloud Recording, otherwise 0 for Phone.
    \item \( D \): 1 for Speaker-side Denoising, otherwise 0.
\end{enumerate}
\noindent
We then formulate the main effects and interactions:
\begin{align}
M = \{1, G, C, D, G \cdot C, G \cdot D, C \cdot D, G \cdot C \cdot D\} .
\end{align} 

Given each acoustic feature and interactions associated with coefficient \( \theta_{i,m} \) where \( i \in [0,25] \) and \( m \in M \), outcomes \( Y_{\text{PESQ}} \) and \( Y_{\text{STOI}} \) are modeled by:
\begin{align}
Y = \sum_{i=0}^{25} \sum_{m \in M} \theta_{i,m} (m \cdot X_i) + \epsilon
\end{align}
with \( \epsilon \) indicating the residual variance, encompassing unexplained variation.

Applying Blinder–Oaxaca decomposition, we unpack the influence of any interaction \(I\) from \(M\), segmenting the total effect for clarity.\cite{oaxaca_decomposition} We employ the notation:
\begin{align}
\Delta \overline{X_i} &= \overline{X_i}_{I=1} - \overline{X_i}_{I=0}, \\
\Delta \overline{\theta_{i,m}} &= \overline{\theta_{i,m}}_{I=1} - \overline{\theta_{i,m}}_{I=0}.
\end{align}

The \textbf{Endowment Effect} is defined as:
\begin{align}
\Delta X_I = \sum_{i} \sum_{m} \Delta \overline{X_i} \overline{\theta_{i,m}}_{I=0}.
\end{align}
This delineates the variance from inherent differences in the states of \(I\), analogized as measuring variations due to signal source alterations.

The \textbf{Coefficient Effect} is expressed as:
\begin{align}
\Delta \theta_I = \sum_{i} \sum_{m} \overline{X_i}_{I=1} \Delta \overline{\theta_{i,m}}.
\end{align}
This elucidates the change in value of certain features depending on \(I\), akin to changes in filter coefficients.

The \textbf{Interaction Effect} is described by:
\begin{align}
\Delta X\Delta \theta_I = \sum_{i} \sum_{m} \Delta \overline{X_i} \Delta \overline{\theta_{i,m}}.
\end{align}
This reveals the compounded impact when both feature values and their coefficients shift together, mirroring simultaneous signal and processing alterations.

Conclusively, the cumulative variation due to \(I\) is:
\begin{align}
\Delta Y_I = \Delta X_I + \Delta \theta_I + \Delta X\Delta \theta_I.
\end{align}
This breakdown offers a deep understanding of the interplay between acoustic features and interactions in diverse telecommunication environments.

\subsection{Decomposition Of STOI On Acoustic Error:}

The analysis of the Short-Time Objective Intelligibility (STOI) metric in relation to acoustic errors reveals fascinating insights. The base effect, which operates as our benchmark, indicates an endowment effect of -0.366, with no variations attributed to coefficient or interaction effects. When examining the Google Meets (G) platform, we witness an improvement, as the collective effect rises to -0.252 due to the coefficient and interaction effects. Conversely, the Cloud usage (C) demonstrates a virtually neutral collective effect, landing at 0.002. In the case of Speaker-side denoising (D), the collective effect closely mirrors the base at -0.361. The interaction effects of Google Meets with Cloud (G\_C) and Google Meets with Denoising (G\_D) exhibit collective effects of -0.119 and -0.409 respectively. The cumulative interaction of Google Meets, Cloud, and Denoising (G\_C\_D) results in a collective effect of -0.269.

\begin{table*}[!ht]\centering
\caption{Comparison between Google Meet and Zoom Platform}\label{tab:comparison}
\setlength{\tabcolsep}{2.2pt}
\renewcommand{\arraystretch}{1.1}
\scriptsize
\begin{tabular}{|l|*{24}{c|}} \hline
& \multicolumn{12}{c|}{\textbf{Google Meet}} & \multicolumn{12}{c|}{\textbf{Zoom}} \\ \cline{2-25}
& \multicolumn{6}{c|}{\textbf{Cloud Recording}} & \multicolumn{6}{c|}{\textbf{Cellular Mobile Recording}} & \multicolumn{6}{c|}{\textbf{Cloud Recording}} & \multicolumn{6}{c|}{\textbf{Cellular Mobile Recording}} \\ \cline{2-25}
& \multicolumn{3}{c|}{\textbf{Sender Denoised}} & \multicolumn{3}{c|}{\textbf{Sender Natural}} & \multicolumn{3}{c|}{\textbf{Sender Denoised}} & \multicolumn{3}{c|}{\textbf{Sender Natural}} & \multicolumn{3}{c|}{\textbf{Sender Denoised}} & \multicolumn{3}{c|}{\textbf{Sender Natural}} & \multicolumn{3}{c|}{\textbf{Sender Denoised}} & \multicolumn{3}{c|}{\textbf{Sender Natural}} \\ \cline{2-25}
& \textbf{\tiny Relay} & \textbf{\tiny FSNet} & \textbf{\tiny Demucs} & \textbf{\tiny Relay} & \textbf{\tiny FSNet} & \textbf{\tiny Demucs} & \textbf{\tiny Relay} & \textbf{\tiny FSNet} & \textbf{\tiny Demucs} & \textbf{\tiny Relay} & \textbf{\tiny FSNet} & \textbf{\tiny Demucs} & \textbf{\tiny Relay} & \textbf{\tiny FSNet} & \textbf{\tiny Demucs} & \textbf{\tiny Relay} & \textbf{\tiny FSNet} & \textbf{\tiny Demucs} & \textbf{\tiny Relay} & \textbf{\tiny FSNet} & \textbf{\tiny Demucs} & \textbf{\tiny Relay} & \textbf{\tiny FSNet} & \textbf{\tiny Demucs} \\ \hline
\textbf{composite\_0} &2.18 &\cellcolor[HTML]{ffc0cb}-.06 &\cellcolor[HTML]{90ee90}+.56 &1.65 &\cellcolor[HTML]{90ee90}+.66 &\cellcolor[HTML]{90ee90}+1.4 &2.16 &\cellcolor[HTML]{ffc0cb}-.58 &\cellcolor[HTML]{ffc0cb}-.10 &1.64 &\cellcolor[HTML]{ffc0cb}-.43 &\cellcolor[HTML]{90ee90}+.15 &2.05 &\cellcolor[HTML]{ffc0cb}-.07 &\cellcolor[HTML]{90ee90}+.53 &1.59 &\cellcolor[HTML]{90ee90}+.50 &\cellcolor[HTML]{90ee90}+1.3 &1.63 &\cellcolor[HTML]{ffc0cb}-.35 &\cellcolor[HTML]{90ee90}+.02 &1.25 &\cellcolor[HTML]{ffc0cb}-.02 &\cellcolor[HTML]{90ee90}+.46 \\ \hline
\textbf{composite\_1} &2.49 &\cellcolor[HTML]{90ee90}+.01 &\cellcolor[HTML]{ffc0cb}-.00 &2.12 &\cellcolor[HTML]{90ee90}+.56 &\cellcolor[HTML]{90ee90}+.51 &2.34 &\cellcolor[HTML]{ffc0cb}-.12 &\cellcolor[HTML]{ffc0cb}-.09 &1.89 &\cellcolor[HTML]{ffc0cb}-.01 &\cellcolor[HTML]{90ee90}+.03 &2.24 &\cellcolor[HTML]{90ee90}+.06 &\cellcolor[HTML]{90ee90}+.05 &1.90 &\cellcolor[HTML]{90ee90}+.53 &\cellcolor[HTML]{90ee90}+.56 &2.06 &\cellcolor[HTML]{ffc0cb}-.17 &\cellcolor[HTML]{ffc0cb}-.07 &1.85 &\cellcolor[HTML]{90ee90}+.01 &\cellcolor[HTML]{90ee90}+.08 \\ \hline
\textbf{composite\_2} &2.21 &\cellcolor[HTML]{ffc0cb}-.01 &\cellcolor[HTML]{90ee90}+.28 &1.61 &\cellcolor[HTML]{90ee90}+.75 &\cellcolor[HTML]{90ee90}+1.0 &2.05 &\cellcolor[HTML]{ffc0cb}-.45 &\cellcolor[HTML]{ffc0cb}-.17 &1.53 &\cellcolor[HTML]{ffc0cb}-.34 &\cellcolor[HTML]{ffc0cb}-.00 &1.97 &\cellcolor[HTML]{90ee90}+.00 &\cellcolor[HTML]{90ee90}+.30 &1.48 &\cellcolor[HTML]{90ee90}+.62 &\cellcolor[HTML]{90ee90}+1.0 &1.61 &\cellcolor[HTML]{ffc0cb}-.37 &\cellcolor[HTML]{ffc0cb}-.07 &1.29 &\cellcolor[HTML]{ffc0cb}-.07 &\cellcolor[HTML]{90ee90}+.20 \\ \hline
\textbf{csii\_0} &0.80 &\cellcolor[HTML]{ffc0cb}-.00 &\cellcolor[HTML]{90ee90}+.00 &0.82 &\cellcolor[HTML]{90ee90}+.04 &\cellcolor[HTML]{90ee90}+.04 &0.67 &\cellcolor[HTML]{ffc0cb}-.01 &\cellcolor[HTML]{ffc0cb}-.00 &0.46 &\cellcolor[HTML]{ffc0cb}-.02 &\cellcolor[HTML]{ffc0cb}-.00 &0.79 &\cellcolor[HTML]{90ee90}+.00 &\cellcolor[HTML]{90ee90}+.00 &0.74 &\cellcolor[HTML]{90ee90}+.04 &\cellcolor[HTML]{90ee90}+.04 &0.50 &\cellcolor[HTML]{ffc0cb}-.03 &\cellcolor[HTML]{ffc0cb}-.00 &0.63 &\cellcolor[HTML]{ffc0cb}-.06 &\cellcolor[HTML]{ffc0cb}-.00 \\ \hline
\textbf{csii\_1} &0.67 &\cellcolor[HTML]{90ee90}+.00 &\cellcolor[HTML]{90ee90}+.00 &0.63 &\cellcolor[HTML]{90ee90}+.08 &\cellcolor[HTML]{90ee90}+.09 &0.55 &\cellcolor[HTML]{ffc0cb}-.03 &\cellcolor[HTML]{ffc0cb}-.01 &0.31 &\cellcolor[HTML]{ffc0cb}-.01 &\cellcolor[HTML]{90ee90}+.00 &0.65 &\cellcolor[HTML]{90ee90}+.00 &\cellcolor[HTML]{90ee90}+.01 &0.58 &\cellcolor[HTML]{90ee90}+.08 &\cellcolor[HTML]{90ee90}+.09 &0.39 &\cellcolor[HTML]{ffc0cb}-.05 &\cellcolor[HTML]{ffc0cb}-.01 &0.47 &\cellcolor[HTML]{ffc0cb}-.05 &\cellcolor[HTML]{ffc0cb}-.01 \\ \hline
\textbf{csii\_2} &0.45 &\cellcolor[HTML]{90ee90}+.00 &\cellcolor[HTML]{90ee90}+.01 &0.30 &\cellcolor[HTML]{90ee90}+.16 &\cellcolor[HTML]{90ee90}+.18 &0.34 &\cellcolor[HTML]{ffc0cb}-.03 &\cellcolor[HTML]{ffc0cb}-.01 &0.09 &\cellcolor[HTML]{90ee90}+.01 &\cellcolor[HTML]{90ee90}+.02 &0.38 &\cellcolor[HTML]{90ee90}+.02 &\cellcolor[HTML]{90ee90}+.02 &0.27 &\cellcolor[HTML]{90ee90}+.14 &\cellcolor[HTML]{90ee90}+.16 &0.17 &\cellcolor[HTML]{ffc0cb}-.04 &\cellcolor[HTML]{ffc0cb}-.00 &0.19 &\cellcolor[HTML]{ffc0cb}-.00 &\cellcolor[HTML]{90ee90}+.02 \\ \hline
\textbf{fwSNRseg} &11.1 &\cellcolor[HTML]{90ee90}+.00 &\cellcolor[HTML]{90ee90}+.12 &9.82 &\cellcolor[HTML]{90ee90}+1.8 &\cellcolor[HTML]{90ee90}+2.2 &7.98 &\cellcolor[HTML]{ffc0cb}-.71 &\cellcolor[HTML]{ffc0cb}-.14 &4.26 &\cellcolor[HTML]{90ee90}+.09 &\cellcolor[HTML]{90ee90}+.67 &10.5 &\cellcolor[HTML]{90ee90}+.06 &\cellcolor[HTML]{90ee90}+.18 &9.45 &\cellcolor[HTML]{90ee90}+1.9 &\cellcolor[HTML]{90ee90}+2.5 &5.63 &\cellcolor[HTML]{ffc0cb}-.67 &\cellcolor[HTML]{90ee90}+.07 &4.80 &\cellcolor[HTML]{ffc0cb}-.04 &\cellcolor[HTML]{90ee90}+.74 \\ \hline
\textbf{llr} &1.59 &\cellcolor[HTML]{90ee90}+.03 &\cellcolor[HTML]{ffc0cb}-.32 &1.64 &\cellcolor[HTML]{ffc0cb}-.05 &\cellcolor[HTML]{ffc0cb}-.60 &1.44 &\cellcolor[HTML]{90ee90}+.18 &\cellcolor[HTML]{ffc0cb}-.00 &1.51 &\cellcolor[HTML]{90ee90}+.17 &\cellcolor[HTML]{ffc0cb}-.11 &1.52 &\cellcolor[HTML]{90ee90}+.06 &\cellcolor[HTML]{ffc0cb}-.26 &1.58 &\cellcolor[HTML]{ffc0cb}-.00 &\cellcolor[HTML]{ffc0cb}-.55 &1.52 &\cellcolor[HTML]{90ee90}+.08 &\cellcolor[HTML]{ffc0cb}-.02 &1.71 &\cellcolor[HTML]{ffc0cb}-.04 &\cellcolor[HTML]{ffc0cb}-.26 \\ \hline
\textbf{ncm} &0.83 &\cellcolor[HTML]{ffc0cb}-.00 &\cellcolor[HTML]{90ee90}+.00 &0.79 &\cellcolor[HTML]{90ee90}+.08 &\cellcolor[HTML]{90ee90}+.09 &0.72 &\cellcolor[HTML]{ffc0cb}-.06 &\cellcolor[HTML]{ffc0cb}-.02 &0.59 &\cellcolor[HTML]{ffc0cb}-.08 &\cellcolor[HTML]{ffc0cb}-.01 &0.88 &\cellcolor[HTML]{90ee90}+.00 &\cellcolor[HTML]{90ee90}+.01 &0.79 &\cellcolor[HTML]{90ee90}+.09 &\cellcolor[HTML]{90ee90}+.10 &0.68 &\cellcolor[HTML]{ffc0cb}-.15 &\cellcolor[HTML]{ffc0cb}-.03 &0.67 &\cellcolor[HTML]{ffc0cb}-.11 &\cellcolor[HTML]{ffc0cb}-.01 \\ \hline
\textbf{pesq} &2.25 &\cellcolor[HTML]{90ee90}+.02 &\cellcolor[HTML]{90ee90}+.00 &1.64 &\cellcolor[HTML]{90ee90}+.79 &\cellcolor[HTML]{90ee90}+.63 &1.98 &\cellcolor[HTML]{ffc0cb}-.28 &\cellcolor[HTML]{ffc0cb}-.24 &1.55 &\cellcolor[HTML]{ffc0cb}-.15 &\cellcolor[HTML]{ffc0cb}-.16 &1.92 &\cellcolor[HTML]{90ee90}+.07 &\cellcolor[HTML]{90ee90}+.07 &1.46 &\cellcolor[HTML]{90ee90}+.68 &\cellcolor[HTML]{90ee90}+.63 &1.70 &\cellcolor[HTML]{ffc0cb}-.35 &\cellcolor[HTML]{ffc0cb}-.18 &1.55 &\cellcolor[HTML]{ffc0cb}-.16 &\cellcolor[HTML]{ffc0cb}-.17 \\ \hline
\textbf{SNRseg} &-0.7 &\cellcolor[HTML]{90ee90}+.08 &\cellcolor[HTML]{90ee90}+.04 &-0.7 &\cellcolor[HTML]{90ee90}+1.6 &\cellcolor[HTML]{90ee90}+2.0 &-0.2 &\cellcolor[HTML]{90ee90}+.40 &\cellcolor[HTML]{90ee90}+.34 &-1.9 &\cellcolor[HTML]{90ee90}+1.0 &\cellcolor[HTML]{90ee90}+1.4 &-1.5 &\cellcolor[HTML]{90ee90}+.23 &\cellcolor[HTML]{90ee90}+.25 &-1.8 &\cellcolor[HTML]{90ee90}+1.9 &\cellcolor[HTML]{90ee90}+2.4 &-1.2 &\cellcolor[HTML]{90ee90}+.34 &\cellcolor[HTML]{90ee90}+.40 &-2.4 &\cellcolor[HTML]{90ee90}+1.4 &\cellcolor[HTML]{90ee90}+1.7 \\ \hline
\textbf{stoi} &0.92 &\cellcolor[HTML]{ffc0cb}-.00 &\cellcolor[HTML]{90ee90}+.00 &0.89 &\cellcolor[HTML]{90ee90}+.03 &\cellcolor[HTML]{90ee90}+.04 &0.88 &\cellcolor[HTML]{ffc0cb}-.04 &\cellcolor[HTML]{ffc0cb}-.02 &0.75 &\cellcolor[HTML]{ffc0cb}-.04 &\cellcolor[HTML]{ffc0cb}-.02 &0.91 &\cellcolor[HTML]{90ee90}+.00 &\cellcolor[HTML]{90ee90}+.00 &0.86 &\cellcolor[HTML]{90ee90}+.04 &\cellcolor[HTML]{90ee90}+.04 &0.81 &\cellcolor[HTML]{ffc0cb}-.08 &\cellcolor[HTML]{ffc0cb}-.02 &0.80 &\cellcolor[HTML]{ffc0cb}-.06 &\cellcolor[HTML]{ffc0cb}-.03 \\ \hline
\textbf{wss} &24.6 &\cellcolor[HTML]{ffc0cb}-.22 &\cellcolor[HTML]{90ee90}+1.2 &35.8 &\cellcolor[HTML]{ffc0cb}-10. &\cellcolor[HTML]{ffc0cb}-11. &31.3 &\cellcolor[HTML]{90ee90}+1.6 &\cellcolor[HTML]{90ee90}+.28 &52.2 &\cellcolor[HTML]{90ee90}+1.6 &\cellcolor[HTML]{ffc0cb}-3.0 &30.9 &\cellcolor[HTML]{ffc0cb}-1.8 &\cellcolor[HTML]{ffc0cb}-.88 &44.3 &\cellcolor[HTML]{ffc0cb}-12. &\cellcolor[HTML]{ffc0cb}-15. &43.9 &\cellcolor[HTML]{90ee90}+4.2 &\cellcolor[HTML]{90ee90}+1.5 &53.0 &\cellcolor[HTML]{ffc0cb}-1.2 &\cellcolor[HTML]{ffc0cb}-7.2 \\
\hline
\end{tabular}
\end{table*}
\begin{table}[!htp]\centering
\scriptsize
\begin{tabular}{|lr|lr|}\hline
\cellcolor[HTML]{ffc0cb} & Negative Change Over Relay  &\cellcolor[HTML]{90ee90} & Positive Change Over Relay \\
\hline
\end{tabular}
\end{table}

\subsection{Decomposition of PESQ On Acoustic Error}

Turning our attention to the Perceptual Evaluation of Speech Quality (PESQ) metric, a profound deviation from the base effect of -1.872 is evident. The Google Meets (G) environment, intriguingly, magnifies this to a steeper -2.432 due to its coefficient effect. Cloud usage (C) pushes the collective effect to -2.181, primarily driven by its coefficient and interaction effects. The Speaker-side denoising (D) effect indicates the most pronounced drop at -2.777, stemming largely from its endowment and coefficient effects. The dual interactions of Google Meets with Cloud (G\_C) and with Denoising (G\_D) lead to collective effects of -2.669 and -3.216, respectively. Lastly, the trilateral interaction (G\_C\_D) reaches the deepest collective effect of -3.382, encapsulating the intricate dynamics of these three parameters in tandem.

In the intricate landscape of VoIP telecommunications, these findings underscore the necessity to delve beyond traditional paradigms. Our analytical foray into the PESQ and STOI metrics unravels the delicate tapestry of interactions that govern the acoustic fidelity in a VoIP setup. By deploying the Oaxaca decomposition, a technique primarily nestled in the precincts of econometrics, we've been able to discern the nuanced contrasts that arise when speech undergoes VoIP transformations. This analytical exercise not only bolsters our grasp over these transformations but also paves the way for future endeavors that seek to refine the acoustic experience in VoIP-mediated communications.

\section{Speech Clarity and Quality Evaluation}

In the context of VoIP systems, quantifying speech clarity and audio fidelity is paramount. Our methodical evaluation using the pysepm evaluation suite \cite{schmiph2_pysepm_2023} provides insights into the objective measures indicative of speech quality and intelligibility in VoIP transmissions \cite{rix2002perceptual} \cite{ma2009objective}. Specific models such as time-domain Demucs \cite{defossez2020real} and time-frequency domain FullSubNet (FSNet) \cite{hao2021fullsubnet} exhibit varying degrees of improvement or degradation, contingent upon the environment. Notably, cloud recordings hint at potential enhancements, whereas cellular scenarios typically indicate a likely deterioration in performance. An intriguing observation is that FullSubNet, when applied to Google Meets without speaker-side denoising, outperforms its counterpart with speaker-side denoising. As the results span a spectrum of outcomes, readers are urged to delve deeper and select metrics that resonate most with their application's requirements \cite{loizou2013speech}, informing integration decisions in VoIP deployment.

\section{Conclusion}

In the rapidly evolving realm of VoIP telecommunications, there exists an acute need for datasets that can capture the true essence and challenges of speech dynamics in this domain. The VoIP-DNS-Tiny dataset introduced and utilized in this study stands as a significant milestone in fulfilling this need. While our innovative approach, leveraging the Oaxaca decomposition technique, demonstrates one possible methodology to examine the intricacies of VoIP-modulated acoustics, the dataset's true potential lies in its relevance to IP use cases.

By providing a comprehensive suite of VoIP samples, complete with variations in denoising settings and receiver types, our dataset offers an invaluable canvas for researchers and technologists to rigorously test, refine, and benchmark their models. The out-of-domain nature of the set especially underscores the importance of real-world context in model evaluation. Before deployment in actual VoIP scenarios, understanding a model's behavior on this dataset can serve as a litmus test for its robustness and reliability.

Looking forward, we encourage the wider academic and industrial communities to harness this dataset's potential. Whether it's to validate existing models or pioneer novel methodologies, VoIP-DNS-Tiny promises to be an instrumental tool. Our future work will diversify broader experimental designs, encompassing varied network configurations, hardware, and global nuances. Through collective endeavors, we aspire to catalyze advancements in VoIP research, paving the way for enhanced user experiences worldwide.



\bibliographystyle{IEEEtran}
\bibliography{mybib}

\begin{thebibliography}{10}
\providecommand{\url}[1]{#1}
\csname url@samestyle\endcsname
\providecommand{\newblock}{\relax}
\providecommand{\bibinfo}[2]{#2}
\providecommand{\BIBentrySTDinterwordspacing}{\spaceskip=0pt\relax}
\providecommand{\BIBentryALTinterwordstretchfactor}{4}
\providecommand{\BIBentryALTinterwordspacing}{\spaceskip=\fontdimen2\font plus
\BIBentryALTinterwordstretchfactor\fontdimen3\font minus \fontdimen4\font\relax}
\providecommand{\BIBforeignlanguage}[2]{{%
\expandafter\ifx\csname l@#1\endcsname\relax
\typeout{** WARNING: IEEEtran.bst: No hyphenation pattern has been}%
\typeout{** loaded for the language `#1'. Using the pattern for}%
\typeout{** the default language instead.}%
\else
\language=\csname l@#1\endcsname
\fi
#2}}
\providecommand{\BIBdecl}{\relax}
\BIBdecl

\bibitem{arora1999voice}
R.~Arora and R.~Jain, ``Voice over ip: Protocols and standards,'' \emph{Network Magazine}, 1999.

\bibitem{bergstra2003itu}
J.~A. Bergstra and C.~A. Middelburg, ``Itu-t recommendation g. 107,'' 2003.

\bibitem{rosenberg2002sip}
J.~Rosenberg, H.~Schulzrinne, and G.~e.~a. Camarillo, ``Sip: session initiation protocol,'' 2002.

\bibitem{bolot1993characterizing}
J.-C. Bolot, ``Characterizing end-to-end packet delay and loss,'' \emph{J. High Speed Networks}, 1993.

\bibitem{fine_grained}
M.~Yang, J.~Konan, D.~Bick, A.~Kumar, S.~Watanabe, and B.~Raj, ``{Improving Speech Enhancement through Fine-Grained Speech Characteristics},'' in \emph{Proc. Interspeech 2022}, 2022, pp. 2953--2957.

\bibitem{taploss}
Y.~Zeng, J.~Konan, S.~Han, D.~Bick, M.~Yang, A.~Kumar, S.~Watanabe, and B.~Raj, ``Taploss: A temporal acoustic parameter loss for speech enhancement,'' 2023.

\bibitem{paaploss}
M.~Yang, J.~Konan, D.~Bick, Y.~Zeng, S.~Han, A.~Kumar, S.~Watanabe, and B.~Raj, ``Paaploss: A phonetic-aligned acoustic parameter loss for speech enhancement,'' 2023.

\bibitem{haykin2008communication}
S.~Haykin, \emph{Communication systems}.\hskip 1em plus 0.5em minus 0.4em\relax John Wiley \& Sons, 2008.

\bibitem{proakis2008digital}
J.~G. Proakis, \emph{Digital communications}.\hskip 1em plus 0.5em minus 0.4em\relax McGraw-Hill, 2008.

\bibitem{konan2023improving}
J.~Konan, O.~Bhargave, and S.~e.~a. Agnihotri, ``Improving perceptual quality, intelligibility, and acoustics on voip,'' \emph{arXiv:2303.09048}, 2023.

\bibitem{davis1980comparison}
S.~Davis and P.~Mermelstein, ``Comparison of parametric representations for monosyllabic word recognition,'' \emph{IEEE Trans. Acoust., Speech, Signal Process.}, 1980.

\bibitem{chhetri2023speech}
S.~Chhetri, M.~S. Joshi, and C.~V. e.~a. Mahamuni, ``Speech enhancement: A survey of approaches and applications,'' in \emph{ICECAA '23}, 2023.

\bibitem{virtanen2012techniques}
T.~Virtanen, R.~Singh, and B.~Raj, \emph{Techniques for noise robustness in automatic speech recognition}.\hskip 1em plus 0.5em minus 0.4em\relax John Wiley \& Sons, 2012.

\bibitem{reddy2020interspeech}
C.~K. A. e.~a. Reddy, ``The interspeech 2020 deep noise suppression challenge: Datasets, subjective testing framework, and challenge results,'' \emph{arXiv preprint arXiv:2005.13981}, 2020.

\bibitem{campbell2015experimental}
D.~Campbell and J.~Stanley, \emph{Experimental and quasi-experimental designs for research}.\hskip 1em plus 0.5em minus 0.4em\relax Ravenio books, 2015.

\bibitem{nuc_user_guide}
\emph{User Guide for NUC10i7FNH, NUC10i5FNH, NUC10i3FNH}, Intel, 2023.

\bibitem{samsung_galaxy_a13_manual}
\emph{Samsung Galaxy A13 5G A136 User Manual}, Samsung, 2023.

\bibitem{mixpre6_ii_manual}
\emph{User manual Sound Devices MixPre-6 II}, Sound Devices, 2023.

\bibitem{william2002voip}
C.~William, ``Voip service quality: measuring and evaluating packet-switched voice,'' \emph{USA: McGraw-Hill Netw. Prof.}, 2002.

\bibitem{rix2002perceptual}
A.~e.~a. Rix, ``Perceptual evaluation of speech quality (pesq) part i--time-delay compensation,'' \emph{J. Audio Eng. Soc.}, 2002.

\bibitem{taal2011algorithm}
C.~e.~a. Taal, ``An algorithm for intelligibility prediction of time--frequency weighted noisy speech,'' \emph{IEEE Trans. Audio, Speech, and Language Processing}, 2011.

\bibitem{oaxaca1973male}
R.~Oaxaca, ``Male-female wage differentials in urban labor markets,'' \emph{Int. Econ. Rev.}, 1973.

\bibitem{blinder1973wage}
A.~S. Blinder, ``Wage discrimination: reduced form and structural estimates,'' \emph{J. Human Res.}, 1973.

\bibitem{flanagan2012voip}
W.~Flanagan, \emph{VoIP and unified communications: internet telephony and the future voice network}.\hskip 1em plus 0.5em minus 0.4em\relax John Wiley \& Sons, 2012.

\bibitem{egemaps}
F.~e.~a. Eyben, ``The geneva minimalistic acoustic parameter set (gemaps) for voice research and affective computing,'' \emph{IEEE transactions on affective computing}, vol.~7, no.~2, pp. 190--202, 2015.

\bibitem{oaxaca_decomposition}
B.~Jann, ``The blinder–oaxaca decomposition for linear regression models,'' \emph{Stata J.}, 2008.

\bibitem{schmiph2_pysepm_2023}
schmiph2, ``pysepm - python speech enhancement performance measures.''

\bibitem{ma2009objective}
J.~e.~a. Ma, ``Objective measures for predicting speech intelligibility in noisy conditions,'' \emph{J. Acoust. Soc. Am.}, 2009.

\bibitem{defossez2020real}
A.~e.~a. Defossez, ``Real time speech enhancement in the waveform domain,'' \emph{arXiv preprint arXiv:2006.12847}, 2020.

\bibitem{hao2021fullsubnet}
X.~e.~a. Hao, ``Fullsubnet: Full-band and sub-band fusion for real-time single-channel speech enhancement,'' in \emph{ICASSP 2021}, 2021.

\bibitem{loizou2013speech}
P.~Loizou, \emph{Speech enhancement: theory and practice}.\hskip 1em plus 0.5em minus 0.4em\relax CRC Press, 2013.

\end{thebibliography}

\end{document}